\documentstyle[12pt,epsfig,psfig]{article}
\begin{document}

\def\ii{\'{\char'20}}
\def\r{\rightarrow}
\def\err{\end{array}}
\def\bea{\begin{eqnarray}}
\def\eea{\end{eqnarray}}
\newcommand{\beq}{\begin{equation}}
\newcommand{\eeq}{\end{equation}}
\newcommand{\nn}{\nonumber}
\def\bp{{\bf p}}
\def\bk{{\bf k}}
\def\bq{{\bf q}}

\begin{center}
{\Large \bf Instanton propagator in scalar model: 
exact expression and contribution to instanton
induced processes} \\

\vspace{4mm}

Yu.A. Kubyshin\footnote{e-mail: \tt kubyshin@theory.npi.msu.su} \\
Institute for Nuclear Physics, Moscow State University  \\
119899 Moscow, Russia \\
and \\
P.G. Tinyakov\footnote{e-mail: \tt peter@ms2.inr.ac.ru} \\
Institute for Nuclear Research of the Russian Academy of Sciences \\
60th October Anniversary prospect, 7a \\
117312 Moscow, Russia
\end{center}

\begin{abstract}
The propagator in the instanton background in the
$(- \lambda \phi^{4})$ scalar model in four dimensions is studied.
Leading and sub-leading terms of its asymptotics for large momenta
and its on-shell double residue are calculated. These results are
applied to the analysis of the initial state and initial-final state
corrections and the calculation of the next-to-leading (propagator)
correction to the exponent of the cross section of
multiparticle scattering processes.
\end{abstract}

\section{Introduction}

In theories with non-trivial structure of vacua 
a number of interesting physical effects, induced by instanton
solutions, appear. 
In the present article we will study shadow processes 
\cite{shadow}. These are non-perturbative processes in which both the
initial and the final state are in the false vacuum.  Apart from
standard perturbative contributions, the processes which start and end
in the false vacuum acquire additional contributions due to the
underbarrier tunelling of the system to another vacuum and its return
to the initial one. This transition is obviously induced by an
instanton solution and goes through the intermediate state containing
a bubble of the true vacuum. We would like to mention that other 
examples of instanton induced processes are transitions with baryon 
number violation between the vacua in the electroweak theory \cite{EWT-vac} 
and the decay of a metastable
(false) vacuum due to underbarrier tunelling from a false vacuum to
the true one \cite{VolKobOkun}.  

Much work has been done to study the instanton induced transitions,
and quite effective techniques for the calculation of the
probabilities of such trunsitions have been developed \cite{Ri1,KRT} 
(see Refs. \cite{Mat,Ti1} for a review). 
We will study the instanton contribution to the total cross section 
$\sigma_{2}(E)$ of a process ($2 \; \rightarrow$ any) with two 
initial particles of the total energy $E$ in the 
$(-\lambda \phi^{4})$-theory. 
There is a number of arguments showing 
that $\sigma_{2}(E)$ can be presented in the following 
exponential form:   
\beq
\sigma_{2}(E) \sim e^{\frac{1}{\lambda} F(\epsilon) +
{\cal O}(1) },
         \label{sigma2}
\eeq
where $\lambda$ is the coupling constant in the model, 
$\epsilon=E/E_{sph}$ and $E_{sph}$ is the energy of the 
sphaleron configuration which
characterizes the height of the barrier separating the vacua.
The leading order approximation of the function $F(\epsilon)$ 
for small $\epsilon$ was studied in Refs. \cite{Ti1,KT}. The
next-to-leading term is a propagator correction for it includes
contributions from the propagator in the instanton background.
Hence, calculation of the next-to-leading correction requires knowledge
of the instanton propagator. It turns out that in the
$(-\lambda \phi^{4})$-theory an exact expression for the
instanton propagator can be obtained. Calculation and discussion of the
propagator correction to the function $F(\epsilon)$ 
is one of the purposes of this article.

An important issue is that of the validity of formula 
(\ref{sigma2}). In the
electroweak theory a proof based on the properties of the propagator
in the instanton background was given in Ref.\cite{Mu92}. We
apply the arguments of Ref. \cite{Mu92} in the $(-\lambda \phi^{4})$-theory
making use of the explicit expression for the propagator.

According to arguments of Refs. \cite{RT,Ti2} for the 
multiparticle initial state the total 
cross section is semiclassical and has the form 
\beq
\sigma_{N}(E) \sim e^{\frac{1}{\lambda} F(\epsilon, \nu) +
{\cal O}(1) },    \label{sigma-N}
\eeq
where $N$ is the number of initial particles, $\nu = N/N_{sph}$, and 
$N_{sph} \sim 1/ \lambda$ is a characteristic number of particles 
contained in the sphaleron. Note that in the regime $\lambda \to 0$ and 
$\nu$ fixed $N \sim \nu / \lambda$ is a large number. 
The function $F(\epsilon,\nu)$ for the 
$(-\lambda \phi^{4})$-theory was calculated numerically for a certain 
range of $\epsilon$ and $\nu$ in Ref. \cite{KT}. In 
Refs. \cite{RT,Ti2,RST} it was argued that  
the leading exponential term of the two-particle cross section can be
calculated from the following formula:
\beq
\lim_{\lambda \rightarrow 0} \lambda \ln \sigma_{2} =
\lim_{\nu \rightarrow 0} F \left( \frac{E}{E_{sph}}, \nu \right). 
\label{conj}
\eeq
In this conjecture it is assumed that the limit $\nu \rightarrow 0$ 
exists. The problem is that the function $F(E/E_{sph}, \nu)$ is known to
contain contributions singular in $\nu$.  In particular, in the 
$(-\lambda \phi^{4})$-theory such
contributions already appear in the
propagator correction. The conjecture basically claims that 
terms singular in $\nu$ cancel each other in the final answer.  
Its validity, of course,
means that the semiclassical form of the two-particle cross section is
indeed given by Eq. (\ref{sigma2}) with $F(E/E_{sph}) = F(E/E_{sph},
0)$. Verification of conjecture (\ref{conj}) in the next-to-leading
order is another purpose of this paper. Note that different arguments
in favor of this conjecture were given in
Refs.\cite{Mueller,RubReb}.

The plan of the article is the following. In Sect. 2 we
describe the model and discuss the propagator in the instanton 
background. Namely, we present the high energy asymptotics of
the propagator and discuss the implementation of Mueller's idea 
in the scalar model. We also discuss the exact expression of the double
residue of the instanton propagator. In Sect. 3 we apply it for the
evaluation of the next-to-leading order (propagator correction)
of the function $F(\epsilon,\nu)$. There we explicitly demonstrate the
appearance of terms singular in $\nu$ for $\nu \rightarrow 0$
and their cancellation in the final result. Sect. 4 contains some 
discussion of the results. In particular, the range of validity of 
the next-to-leading order approximation is estimated. 

\section{Instanton propagator in the scalar model}

We consider the model of one component real scalar field, defined
by the Minkowskian action
\beq
S = \int d^{4}x \left[ \frac{1}{2} \left( \partial_{\mu} \phi
\right)^{2} - \frac{1}{2} m^{2} \phi^{2} + \frac{\lambda}{4!}
\phi^{4} \right],    \label{action}
\eeq
where $\lambda > 0$. The potential of the model is unbounded from
below, hence the minimum $\phi = 0$ is metastable. Underbarrier
tunelling from this vacuum to the
instability region and its return to the trivial vacuum is the
transition which gives rise to the shadow process we are going
to study here.

Let us consider first the case $m=0$. There is a well known instanton
solution in the massless theory given by the formula
\cite{Fu,Lip}
\beq
\phi_{inst}(x; x_{0}, \rho) = \frac{4\sqrt{3}}{\sqrt{\lambda}}
\frac{\rho}{(x-x_{0})^{2} + \rho^{2}}.   \label{inst}
\eeq
Here $x_{0 \mu}$ is the center of the instanton and $\rho$ is its size. 
Due to the conformal invariance of the massless theory the action of the
instanton does not depend on $\rho$, 
\beq
S_{inst}^{(0)} \equiv S(\phi_{inst}) = \frac{16 \pi^{2}}{\lambda}.
\label{S-inst}
\eeq

In the case $m \neq 0$ the mass term breaks the conformal invariance.
Using standard scaling arguments it can be shown that there are
no regular solutions of the Euclidean equations of motion with finite
action. The decay of the vacuum $\phi=0$ is dominated by the 
constrained instanton, a configuration which can be regarded
as an approximate solution of the equations of motion. It minimizes
the action under the constraint that the size of the configuration
is $\rho$. A formalism for construction of such configurations
and evaluation of the functional  integral was developed in
\cite{Aff}.

When $\rho m \ll 1$ the constrained instanton configuration behaves
like the instanton (\ref{inst}) of the massless theory at
$x \ll \rho$ and as a solution of the free massive theory 
for $x > m^{-1}$. The action of such configuration is
\beq
S_{inst}(\rho) = \frac{16 \pi^{2}}{\lambda} - \frac{24 \pi^{2}}{\lambda}
(\rho m)^{2} \left[ \ln \frac{\rho^{2} m^{2}}{4} + 2 C_{E} + 1 \right]
+ {\cal O} (\rho^{4}m^{4} ),   \label{S-ci}
\eeq
where $C_{E} = 0.577 \ldots$ is the Euler constant. For the class
of constraints mentioned above the terms given in (\ref{S-ci}) do
not depend on the explicit form of the constraint, whereas the
corection ${\cal O}(\rho^{4} m^{4})$  does. In
our analysis we limit ourselves to the constraint independent order
of the approximation.

For $m \neq 0$ the potential barrier separating the trivial vacuum
$\phi =0$ from the instability region is finite. Its height is
characterized by a sphaleron solution, a static $SO(3)$-symmetric
configuration satisfying the equation of motion. In Ref. \cite{KT}
it was found that the sphaleron energy and the sphaleron number of 
particles are 
\beq
E_{sph} = \kappa \frac{m}{\lambda}, \; \; \; \kappa = 113.4 
\; \; \; \mbox{and} \; \; \; N_{sph} = 63 \frac{1}{\lambda}
\label{Esph}
\eeq
respectively. 

Now let us study the propagator in the instanton background. 
It is defined by the operator $\hat{D}_{x}$
of quadratic fluctuations appearing in the expansion of the
action around the instanton solution.
In the massless case this operator is equal to
\[
\hat{D}_{x} = - \frac{\partial^{2}}{\partial x_{\mu}^{2}} +
\frac{\lambda}{2} \phi_{inst}^{2}(x;0,\rho) =
- \frac{\partial^{2}}{\partial x_{\mu}^{2}} + \frac{24 \rho^{2}}
{(\rho^{2} + x^{2})^{2}}.
\]
It can be easily seen that it possesses five zero modes
$\psi_{A}(x)$ $(A=1,2,3,4,5)$ corresponding to the translational
invariance and the scale invariance of the massless theory.
The zero modes can be obtained by differentiation
of the instanton solution with respect to the parameters $x_{0}$
and $\rho$: 
\beq
\psi_{A} \sim \frac{\partial}{\partial \zeta_{A}}
\phi_{inst}(x; x_{0}, \rho)|_{x_{0}=0}, \; \; \;
\zeta_{\mu} = (x_{0})_{\mu}, \; \; \zeta_{5}=\rho.
\label{0-modes}
\eeq
Because of the existence of the zero modes there is an ambiguity 
in the definition of the propagator that can be fixed by imposing 
additional constraints. 
Let $G_{f}(x,y)$ be the inverse of $\hat{D}_{x}$ on the subspace 
of functions orthogonal to some functions $f_{A}(x)$. The latter 
satisfy the only condition that the matrix
\[
\Omega_{AB} = \int dx \psi_{A} (x) f_{B}(x)   
\]
is invertible \cite{LevYaf}. According to this definition 
the instanton propagator satisfies the equation
\beq
\hat{D}_{x} G_{f}(x,y) = \delta(x-y) - \sum_{A} f_{A}(x) \Omega^{-1}_{AB}
\psi_{B}(x)   \label{prop-eqn}
\eeq
and the orthogonality constraints
\beq
\int dx f_{A}(x) G_{f}(x,y) = 0 = \int dy G_{f}(x,y) f_{B}(y).
\label{G-constr}
\eeq
The r.h.s. of Eq. (\ref{prop-eqn}) is the projector onto the subspace
orthogonal to the functions $f_{A}(x)$. Physical results, of course, do
not depend on particular choice of the functions $f_{A}(x)$. 
Below, following the ideas
of Refs.~\cite{MMcLY,Mu92}, we will use the freedom of choosing
constraints (\ref{G-constr}) to eliminate the leading asymptotics of
the instanton propagator and simplify the analysis of the
initial-state corrections.

For a particularly simple and natural choice of
the functions $f_{A}(x)$, namely for 
\beq
f_{A}(x) = w(x) \psi_{A}(x),    \label{f-1}
\eeq
where the weight function
\[
w(x) = \frac{4 \rho^{2}}{(\rho^{2} + x^{2})^{2}}, 
\]
the propagator in the instanton background was calculated explicitly 
in Ref. \cite{Ku} (see also \cite{KubTi}). Allowing some abuse of notation 
we denote this propagator by $G_{\psi}(x,y)$. It is equal to 
\bea
G_{\psi}(x,y) & = & \frac{1}{2\pi^{2}} \frac{\rho^{2}}
{(\rho^{2} + x^{2}) (\rho^{2} + y^{2})}
\left\{ \frac{1}{2d(x,y)} - 3 \ln d(x,y) 
 \right.   \nonumber \\
& - & \left.  \frac{43}{5} + 6 d(x,y) \ln d(x,y) +
\frac{56}{5} d(x,y)  \right\},  \label{propag}
\eea
where 
\[
d(x,y) = \frac{\rho^{2}(x-y)^{2}}{(\rho^{2} + x^{2})
(\rho^{2} + y^{2})}.    
\]

The relation between $G_{\psi}(x,y)$ and the propagator $G_{f}(x,y)$ 
for an arbitrary constraint (\ref{G-constr}) is given by the following
formula: 
\bea
G_{f}(x,y) & = & G_{\psi}(x,y) -
\left( \int dz G_{\psi}(x,z) f_{A}(z) \right) \Omega_{AB}^{-1}
\psi_{B}(y) \nonumber \\
 & - & \psi_{A}(x) \left( \Omega_{AB}^{T} \right)^{-1}
\int dz f_{B}(z) G{\psi}(z,y)  \nonumber \\
& + & \psi_{A}(x) \left( \Omega_{AB}^{T} \right)^{-1}
\left( \int dz dz' f_{B}(z) G{\psi}(z,z') f_{C}(z') \right)
\Omega_{CD}^{-1}  \psi_{D}(y).    \label{G-G}
\eea

The Fourier transform of the instanton propagator is defined
in the standard way:
\[
G_{f}(p,q) = \int dx dy e^{ipx + iqy} G_{f}(x,y).   
\]
In principle, using the exact result, Eq. (\ref{propag}), the function 
$G_{\psi}(p,q)$ can be obtained by direct calculation. 
We did not find the complete expression. Instead we
derived the asymptotic formula for the Fourier transform of the
instanton propagator in the regime when $p^{2}$, $q^{2}$ are
fixed and $s \equiv (p+q)^{2} \rightarrow \infty$. The growing terms
of the asymptotics are given by
\beq
G_{\psi}(p,q) = \frac{16 \pi^{2}}{p^{2} q^{2}} \left[
s\rho^{2} \ln (s \rho^{2}) \Pi_{1}(p,q) +
(s \rho^{2}) \Pi_{2}(p,q) +
\ln (s \rho^{2}) \Pi_{3}(p,q) + \ldots \right],   \label{G-asymp}
\eeq
where
\bea
\Pi_{1}(p,q) & = &  \frac{3}{4}  {\cal S}_{1}(p \rho)
{\cal S}_{1}(q \rho), \label{Pi1} \\
\Pi_{2}(p,q) & = & \frac{3}{2} \left( C_{E} - \frac{1}{15}
- \ln 2 \right) {\cal S}_{1}(p \rho)
{\cal S}_{1}(q \rho), \label{Pi2} \\
\Pi_{3}(p,q) & = & \left\{ {\cal S}_{1}(p\rho)
\left[ \frac{9}{2} {\cal S}_{2} (q\rho) - \left( \frac{27}{4} +
\frac{3}{4} q^{2}\rho^{2} \right) {\cal S}_{1}(q\rho) \right] \right.
\nonumber \\
& + &  \left.
\left[ \frac{9}{2} {\cal S}_{2} (p\rho) - \left( \frac{27}{4} +
\frac{3}{4} p^{2}\rho^{2} \right) {\cal S}_{1}(p\rho) \right]
{\cal S}_{1}(q\rho) - \frac{3}{2}
{\cal S}_{2}(p\rho) {\cal S}_{2}(q\rho) \right\}.  \label{Pi3}
\eea
Here ${\cal S}_{n}(z)$ is defined by 
${\cal S}_{n}(z)= z^{n} K_{n}(z)$, where $K_{n}(z)$ is the 
modified Bessel function. Using the explicit expressions for 
the translational zero modes (see Eq. (\ref{0-modes})) 
normalized with respect to the weight function $w(x)$, 
the first two terms of the asymptotic expansion (\ref{G-asymp}) 
can be written as
\beq
G_{\psi}(p,q) = -\frac{1}{5\rho^{2}} \ln (\rho^{2} s)
\psi_{\mu}(p) \psi_{\mu}(q)
 - \frac{2}{5 \rho^{2}} \left( C_{E} - \frac{1}{15}-\ln 2 \right)
\psi_{\mu}(p) \psi_{\mu}(q) + \ldots \label{G-asymp1}
\eeq
The leading term of the asymptotics of the propagator
in the instanton background was calculated in Ref. \cite{Vo1}.
This result is in complete agreement with the first term
in Eq. (\ref{G-asymp1}).

In Ref. \cite{Mu92} Mueller proposed an idea to use the ambiguity
in the choice of the function $f_{A}(x)$ in order 
to cancel the two leading terms in the asymptotics of the
propagator $G_{\psi}(p,q)$. Then the propagator contribution, 
as well as loop contributions of the initial state corrections disappear. 
As a consequence, such corrections do not exponentiate, i.e., 
do not give contributions to the function $F(\epsilon)$. In addition, 
in this case the initial-final state corrections can be described 
semiclassically. Namely, the effect of the initial state 
lines can be taken into
account by substituting the instanton by a new field
configuration which is a solution
to the classical equation of motion
with an external source (see Ref. \cite{Mu92} for details).

Now we explain how the functions $f_{A}(x)$ can be chosen to 
provide vanishing of the two leading terms of the asymptotics of 
$G_{f}(p,q)$. For this we repeat the arguments of 
Ref. \cite{Mu92}. It turns out that for such functions 
the corresponding propagator constraint (\ref{G-constr}) 
is not relativistically covariant.
Let $p_{1}$ and $p_{2}$ be the arguments of the Fourier
transform of the propagator. We choose a coordinate system such that 
$p_{1 j}=p_{2 j}=0$ for $j=2,3$, whereas $p_{1+} \rho = p_{2}\rho \gg 1$
and $p_{1}^{2}$ and $p_{2}^{2}$ are fixed. Here the $\pm$ components
of the momenta are defined by
\[
p_{j\pm} = \frac{(p_{j})_{0}\pm(p_{j})_{1}}{\sqrt{2}}.
\]
Then $(p_{1},p_{2})\rho^{2} \approx p_{1+} p_{2-} \rho^{2} \gg 1$.
Only the components $f_{\mu}(x)$, corresponding to translations,
modify the leading asymptotics of the propagator.
The Fourier transforms $\tilde{f}_{\mu}(p)$ of the functions
$f_{\mu}(x)$, defining the required propagator constraint, are
chosen in the following way: 
\beq
\tilde{f}_{\mu} (p) = \delta (p_{+}-M) \delta (p_{-}+M)
\bar{f}_{\mu} (p_{2},p_{3}),
\eeq
where $M$ is an arbitrary parameter of the dimension of mass.
Substituting these functions into Eq. (\ref{G-G}) one finds 
after some calculations that
\beq
G_{f}(p_{1},p_{2}) = \psi_{\mu}(p_{1}) {\cal G}_{\mu \nu}
\psi_{\nu} (p_{2}),
\eeq
where the $4 \times 4$ constant
matrix ${\cal G}_{\mu \nu}$ is equal to
\[
{\cal G}_{\mu \nu} = -\frac{1}{5 \rho^{2}} \Omega_{\mu \sigma}^{-1}
 \frac{1}{(2\pi)^{8}} \int d^{4}q_{1} d^{4}q_{2}
f_{\sigma}(-q_{1}) \psi_{\rho}(q_{1})
\ln \frac{(q_{1},q_{2})}{M^{2}} \psi_{\rho}(q_{2}) f_{\tau}(-q_{2})
\left( \Omega^{T} \right)_{\tau \nu}^{-1}.
\]
Using the freedom of choosing the functions $\bar{f}_{\mu}$
one can make the constant real symmetric matrix
${\cal G}_{\mu \nu}$ equal to zero. We would like to stress 
that the knowledge of the exact formulas for the leading 
terms of the asymptotics of $G_{\psi}(p,q)$, Eqs. (\ref{G-asymp}) - 
(\ref{G-asymp1}), allows us to get the explicit expression of the 
matrix ${\cal G}_{\mu \nu}$. This is in contrast with the case of the 
electroweak theory where only a general structure of the analogous 
matrix can be derived \cite{Mu92}. 

For the perturbative calculations of the function
$F(\epsilon,\nu)$ the on-mass-shell residue of the instanton
solution will be needed. By definition it is equal to
\beq
R_{inst}({\bf k}) = \left. (k^{2}+m^{2})
\tilde{\phi}_{inst}(k;0,\rho)\right|_{k_{0}=i\omega_{\bk}},    
\label{R-inst-def}
\eeq 
where $\tilde{\phi}_{inst}(k;x_{0},\rho)$ is the Fourier 
transform of the instanton,
\[
 \tilde{\phi}_{inst}(k;x_{0},\rho) = \int d^{4}x e^{ikx} 
\phi_{inst}(x;x_{0},\rho)
\]
and $\omega_{\bk}=\sqrt{\bk^{2} + m^{2}}$. 
For the instanton solution (\ref{inst}) in the massless theory 
\beq
  R_{inst} = \frac{1}{\sqrt{\lambda}} 16 \sqrt{3}
  \pi^{2} \rho.  \label{R-inst}
\eeq

Correspondingly, to calculate of the next-to-leading correction
to the function $F(\epsilon, \nu)$ we need the expressions
for the double on-mass-shell residues of the instanton propagator.
We will consider the propagator orthogonal to functions (\ref{f-1}). 
Let us introduce the following notations:  
\bea
R_{aa}(\bk,\bq) & = & 
(k^{2} + m^{2}) (q^{2} + m^{2}) \left. G_{\psi}(k_{0},\bk;q_{0},\bq) 
\right|_{k_{0}=i\omega_{\bk},q_{0}=i\omega_{\bq}},  \label{R-def-aa} \\
R_{ab}(\bk,\bq) & = & 
(k^{2} + m^{2}) (q^{2} + m^{2}) \left. G_{\psi}(k_{0},\bk;q_{0},-\bq) 
\right|_{k_{0}=i\omega_{\bk},q_{0}=-i\omega_{\bq}},  \label{R-def-ab} \\
R_{bb}(\bk,\bq) & = & 
(k^{2} + m^{2}) (q^{2} + m^{2}) \left. G_{\psi}(k_{0},-\bk;q_{0},-\bq) 
\right|_{k_{0}=-i\omega_{\bk},q_{0}=-i\omega_{\bq}},  \label{R-def-bb} 
\eea
The indices $a$ and $b$ correspond to initial and final particles, 
respectively (in the notations of Ref. \cite{Ti2}). For the 
scalar massless field $\omega_{\bk}=|\bk|$ and all three residues 
(\ref{R-def-aa})-(\ref{R-def-bb}) can be expressed in terms of one 
function:
\beq
R_{\#}(\bk,\bq) = \rho^{2} R \left(\rho^{2} s^{(0)}_{\#}(\bk,\bq) \right), 
\label{RR}
\eeq
where $\# = aa$, $ab$, $bb$ and the function $s^{(0)}_{\#}(\bk,\bq)$ is 
the $s$-variable for the corresponding particles on the mass shell,
\[
s^{(0)}_{aa}(\bk, \bq) = s^{(0)}_{bb}(\bk, \bq) = 
- s^{(0)}_{ab}(\bk, \bq) = 
- 2(|\bk| |\bq| - \bk\bq ).  
\]
However, in the calculation of the 
next-to-leading order corrections due to non-zero mass  
must be taken into account. It turns out that within the accuracy set by 
Eq. (\ref{S-ci}) it is enough to consider the residues defined through 
the relation 
\beq
R_{\#}(\bk,\bq) = \rho^{2} R \left(\rho^{2} s_{\#}(\bk,\bq) \right), 
\label{RR1}
\eeq
(cf. (\ref{RR})), where the function is calculated for the instanton 
propagator of the {\it massless} theory, whereas the $s$-variable 
is taken for the {\it massive} one:
\bea
s_{aa}(\bk, \bq) & = & s_{bb}(\bk, \bq) =
- 2 m^{2} - 2(\omega_{\bk} \omega_{\bq} - \bk\bq ), \label{saa} \\
s_{ab}(\bk, \bq) & = & 
- 2 m^{2} + 2(\omega_{\bk} \omega_{\bq} - \bk\bq ). \label{sab}
\eea
The consistency of this procedure is discussed in Sect. 4. 
 
The {\it exact} expression for the function $R(s)$ 
was obtained in Ref. \cite{KubTi} and is given by
\bea
R(s) & = & 16 \pi^{2} \left\{ \alpha_{1} \left[ s \ln \frac{s}{4} + 
2 \left(C_{E} - \frac{1}{15} \right) s \right] 
- \alpha_{2} \left[ \ln \frac{s}{4} + 
2 \left( C_{E} + \frac{43}{30} \right) \right] \right\},  \label{R} \\
\alpha_{1} & = & 3/4, \; \; \; \alpha_{2}=3/2.   \nonumber 
\eea
In the next section this result will be used for the calculation
of the next-to-leading correction to the function
$F(\epsilon,\nu)$.

\section{Multiparticle cross section}

Formula (\ref{sigma-N}) for the multiparticle cross-section of
shadow processes comes from the following expression derived in
Ref. \cite{Ti2},
\bea
\sigma_{N} (E) & \sim & \int d^{4}x_{0} d \rho d^{4}\xi d \theta
\exp \left[ - 2 S_{inst}(\rho) +
\frac{1}{\lambda} W^{(1)}(x_{0}, \rho, \xi, \theta) \right. \nonumber \\
& + & \left.
\frac{1}{\lambda} W^{(2)}(x_{0}, \rho, \xi, \theta) + \ldots \right],
\label{sigma-W}
\eea
where we integrate over the position $x_{0}$ and the size $\rho$ of the
instanton, as well as over auxiliary variables $\xi_{\mu}$ and
$\theta$.  We also indicated explicitly the dependence of the action
on the size of the instanton (see Eq. (\ref{S-ci})). The
terms $W^{(i)}$ account for fluctuations in the instanton background:
$W^{(1)}$ corresponds to leading diagrams without propagator
lines, $W^{(2)}$ corresponds to diagrams with one internal
propagator in the instanton background, etc. Diagrams with loops do
not appear in the $(1/\lambda)$ order of the semi-classical
approximation, they contribute to ${\cal O} (1)$ terms in
(\ref{sigma-N}).

General expressions for the functions $W^{(1)}$ and $W^{(2)}$ were 
derived in \cite{Ti2}.
The integrals in Eq. (\ref{sigma-W}) are evaluated by
the saddle point method.
It can be checked that up to the next-to-leading
order the saddle point values of $x_{0}$, $\rho$, $\xi$ and $\theta$
are determined by the leading-order equations. These equations are
obtained by differentiation of the expression
$(-2S_{inst}(\rho) + W^{(1)}/\lambda)$ with respect to
$x_{0}$, $\rho$, $\xi$ and $\theta$. The physically relevant saddle
point has $(x_{0})_i=0$, $\xi_i=0$ ($i=1,2,3$), while $(x_{0})_{0}$, 
$\xi_{0}$ and $\theta$ are purely imaginary. It is convenient to 
introduce the following notations: $x_{0}= i\tau$, $\xi_{0} = i\chi$, 
and $\theta = -i \ln \gamma$. 
In accordance with Eq. (\ref{sigma-W}) the function
$F(\epsilon,\nu)$ is represented as
\beq
F(\epsilon,\nu) = -32 \pi^{2} + F^{(1)}(\epsilon,\nu) +
F^{(2)}(\epsilon,\nu) + \ldots.   \label{F-def}
\eeq
The first term in the r.h.s. is just $(-2\lambda S_{inst}^{(0)})$,
where $S_{inst}^{(0)}$ is the instanton action in the
massless theory, Eq. (\ref{S-inst}). The non-trivial leading
order correction $F^{(1)}(\epsilon,\nu)$ corresponds to the
contribution of
\[
2 \lambda (S_{inst}^{(0)}-S_{inst}(\rho)) + W^{(1)}
\]
in Eq. (\ref{sigma-W}). The next-to-leading (propagator)
correction  $F^{(2)}(\epsilon,\nu)$ is given by $W^{(2)}$ 
evaluated at the saddle point solution.

\subsection{Leading order correction}

For general values of $\epsilon$ and $\nu$ the system of
saddle point equations is too complicated and we studied it numerically.
The results are described at the end of the section.

In the limit of small $\nu$ the calculations simplify considerably. Keeping 
only relevant terms we obtain that  
the function $W^{(1)}$ of Eq. (\ref{sigma-W}) reads 
\bea
 & & \frac{1}{\lambda} W^{(1)}(\tau, \rho, \chi, \gamma) =
 E \chi - N \ln \gamma + 
 \frac{1}{(2\pi)^{3}} \int \frac{d\bp}{2 \omega_{\bp}} 
 R_{inst}(\bp) e^{-\omega_{\bp} \tau} R_{inst}(\bp) \nonumber \\   
 & & + \frac{\gamma}{(2\pi)^{3}} \int \frac{d\bp}{2 \omega_{\bp}} 
 R_{inst}(\bp) e^{-\omega_{\bp} (\chi-\tau)} R_{inst}(\bp) + 
\ldots \nonumber \\
& & = E \chi - N \ln \gamma + 192 \pi^{2} \frac{\rho^{2}m^{2}}{\lambda} 
\left[ \Phi (m\tau) + \frac{\gamma}{m^{2}(\chi - \tau)^{2}} \right] 
+ \ldots ,  \label{W1} 
\eea
where $R_{inst}$ is given by expression
(\ref{R-inst}), $\omega_{\bp}=\sqrt{\bp^{2} + m^{2}}$, and 
\[
\Phi(z) \equiv \frac{1}{z} K_{1}(z). 
\] 
We would like to stress that in these calculations the expression for 
the energy $\omega_{\bp}$ of the massive theory is used, 
whereas it is enough
to substitute the residue $R_{inst}$ of the instanton solution
of the massless theory. This is consistent with the approximation
we are considering in the present paper. The question of validity 
of this procedure is discussed in Sect. 4. 

For further calculations it is convenient to introduce the
variables $\tilde{\epsilon} = E \lambda / m$ and $\tilde{\nu} = 
N \lambda$. From Eqs. (\ref{Esph}) it follows that
$\tilde{\epsilon} = \kappa \epsilon$ and $\tilde{\nu} = 63 \nu$. 
To the leading order in $\nu$ the saddle point solution can be written 
in the form
\beq
\tilde{\rho}^{2} = - \frac{1}{192 \pi^{2} m^{2}}
\frac{\tilde{\epsilon}}{\Phi'(m\tilde{\tau})}; \quad
\tilde{\gamma} = - 4 \left( \frac{\tilde{\nu}}{\tilde{\epsilon}} \right)^{3}
\Phi' (m\tilde{\tau}); \quad
\tilde{\chi} = \tilde{\tau} + \frac{2}{m}
\frac{\tilde{\nu}}{\tilde{\epsilon}}.  \label{saddle}
\eeq
Here the prime denotes the derivative, $\ln C = - \ln 4 + 2 C_{E} + 1$, 
and $\tilde{\tau}=\tilde{\tau}(\epsilon)$
is determined by the equation
\beq
\ln \left( - \frac{\tilde{\epsilon} C e}{192 \pi^{2} \Phi' (m\tilde{\tau})}
  \right) + 4 \Phi (\tilde{m\tau})  = 0.   \label{tau-eqn}
\eeq
Note that the saddle point solution satisfies the relation
\beq
\frac{2 \tilde{\gamma}}{m^{3} (\tilde{\chi} - \tilde{\tau})^{3}} +
\Phi' (m \tilde{\tau}) = 0,     \label{saddle-eql}
\eeq
which will be used later.

Substituting the saddle point solution into Eq. (\ref{W1}) 
we obtain the leading order contribution $F^{(1)}$:
\beq
F^{(1)}(\epsilon,\nu) = \kappa \epsilon \left[ m \tilde{\tau}_{0}
(\epsilon) + \frac{1}{4 \Phi'(m\tilde{\tau}_{0}(\epsilon))} \right]
+{\cal O}(\nu).              \label{F1-1}
\eeq

In the limit $\epsilon \rightarrow 0$ Eq. (\ref{tau-eqn}) 
can be solved iteratively. One gets
\beq
  m \tilde{\tau}(\epsilon)  = \frac{2}{\sqrt{\ln \frac{1}{\epsilon}}}
  + \frac{\ln \ln \frac{1}{\epsilon}}
  {\left( \ln \frac{1}{\epsilon} \right)^{3/2}} + \ldots
\label{tau-1}
\eeq
Then in the leading order in energy solutions (\ref{saddle}) become
\beq
(m \tilde{\rho})^{2} = \frac{1}{48 \pi^{2}}
\frac{\tilde{\epsilon}}{\left( 
\ln \frac{1}{\tilde{\epsilon}} \right)^{3/2}}; \quad
\tilde{\gamma} = \left( \frac{\tilde{\nu}}{\tilde{\epsilon}} \right)^{3}
\left( \ln \frac{1}{\tilde{\epsilon}} \right)^{3/2}; \quad
m(\tilde{\chi} - \tilde{\tau}) = 
2 \frac{\tilde{\nu}}{\tilde{\epsilon}}.  \label{saddle-1}
\eeq
In this regime the function $F^{(1)}(\epsilon, \nu)$ is equal to
\beq
F^{(1)}(\epsilon, \nu) = 2 \frac{\kappa \epsilon}
{\sqrt{ \ln \frac{1}{\epsilon}}} \left[ 1 + {\cal O}
\left( \frac{\ln \ln \frac{1}{\epsilon}}{\ln \frac{1}{\epsilon}}
\right) \right] + {\cal O}(\nu).    \label{F1}
\eeq

\subsection{Propagator correction}

The next-to-leading order function $W^{(2)}$ can be written
as the sum of contributions involving the propagator between final
states, between initial and final states and between initial states, 
respectively:
\beq
W^{(2)} =  W^{(2)}_{(f-f)} + W^{(2)}_{(i-f)} + W^{(2)}_{(i-i)}. 
\label{W2}
\eeq
As we have already mentioned the expressions for these terms are 
given in Ref. \cite{Ti2}. The complete propagator correction was 
calculated numerically, the results are discussed in Sect. 3.3. Here 
we study the propagator correction analytically in the limit 
of small $\nu$. Keeping only relevant contributions we obtain that 
\bea
\frac{1}{\lambda} W^{(2)}_{(f-f)} & = & I_{bb} (\tau, \tau) + \ldots , 
      \label{W2-ff} \\
\frac{1}{\lambda} W^{(2)}_{(i-f)} & = & 
2 \gamma I_{ab} (\tau, \chi - \tau) + \ldots,   \label{W2-if} \\
\frac{1}{\lambda} W^{(2)}_{(i-i)} & = & 
\gamma^{2} I_{aa} (\chi-\tau, \chi-\tau) + \ldots, 
      \label{W2-ii} 
\eea
where
\bea
I_{\#}(\tau_{1}, \tau_{2}) & = & 
\frac{1}{(2\pi)^{6}m^{2}} \int \frac{d \bk}{2\omega_{\bk}}
\frac{d \bq}{2\omega_{\bq}} R_{inst}(\bk) 
e^{-\omega_{\bk}\tau_{1}} R_{\#} (\bk,\bq) 
R_{inst}(\bq) e^{-\omega_{\bq}\tau_{2}} \label{I-def} \\
& = & 48 \frac{\rho^{4}}{\lambda}
\int \frac{d \bk}{2\omega_{\bk}} \frac{d \bq}{2\omega_{\bq}} 
e^{-\omega_{\bk}\tau_{1}} 
\frac{R \left(\rho^{2} s_{\#}(\bk,\bq) \right)}{16 \pi^{2}} 
e^{-\omega_{\bq}\tau_{2}}.  \nonumber 
\eea
The functions $R_{\#}(\bk,\bq)$ and $R(\rho^{2}s)$ are given by Eqs. 
(\ref{RR1}) - (\ref{R}), all necessary 
notations were introduced in Sect. 2. 

In the limit of small $\nu$ the expression for the propagator 
correction in terms of simple integrals can be obtained. 
However, it is quite cumbersome and we
do not present this result here. Instead we calculate
and analyze groups of terms which are singular in $\nu$.
{}From Eqs. (\ref{saddle}) it follows that for the saddle
point solution in the limit $\nu \rightarrow 0$ we have
\beq
m(\tilde{\chi} - \tilde{\tau}) \sim \nu \rightarrow 0, \; \; \;
\tilde{\gamma} \sim \nu^{3}. \label{gxt}
\eeq
Using these properties it is easy to select and calculate the
terms in Eqs. (\ref{W2-if}) and (\ref{W2-ii}) which are singular
in $\nu$. Evaluating these terms at the saddle point solution 
(\ref{saddle}), we obtain that 
\bea
F^{(2)}_{(i-i)} & = & - 32 (192 \pi^{2}) \alpha_{1} \tilde{\rho}^{6}
\frac{\tilde{\gamma}^{2}}{(\tilde{\chi}-\tilde{\tau})^{6}}
\left[ \ln \frac{\tilde{\rho}^{2}}{(\tilde{\chi}-\tilde{\tau})^{2}}
+ \ldots \right]   \nonumber \\
& = &
\frac{8 \alpha_{1}}{(192 \pi)^{2}} \frac{\tilde{\epsilon}^{3}}
{\Phi'(m\tilde{\tau})} \left[ 2 \ln \frac{1}{\tilde{\nu}} +
\ln \left(-\frac{\tilde{\epsilon}^{3}}{768 \pi^{2}\Phi'(m\tilde{\tau})}
\right) + \ldots \right], \label{Fii-1}  \\
F^{(2)}_{(i-f)} & = & 16 (192 \pi^{2}) \alpha_{1} \tilde{\rho}^{6}
\frac{2\tilde{\gamma}}{(\tilde{\chi}-\tilde{\tau})^{3}}
\left[ - \Phi'(m\tilde{\tau}_{0})
\ln \frac{\tilde{\rho}^{2}}{\tilde{\tau}
(\tilde{\chi}-\tilde{\tau})} + \ldots \right]   \nonumber \\
& =  &
- \frac{16 \alpha_{1}}{(192 \pi)^{2}} \frac{\tilde{\epsilon}^{3}}
{\Phi'(m\tilde{\tau})} \left[ \ln \frac{1}{\tilde{\nu}} +
\ln \left(-\frac{\tilde{\epsilon}^{2}}{384 \pi^{2}
m\tilde{\tau} \Phi'(m\tilde{\tau})}\right)  + \ldots \right],
\label{Fif-1}
\eea
where the dots stand for non-singular terms. Summing contributions 
(\ref{Fii-1}) and (\ref{Fif-1}) one gets  
\bea
F^{(2)}_{(i-i)} & + & F^{(2)}_{(i-f)}  \nonumber \\
& = & - 16 (192 \pi^{2}) \alpha_{1} \tilde{\rho}^{6} 
\left[ \frac{\tilde{\gamma}}
{(\tilde{\chi}-\tilde{\tau})^{3}} \left(
\frac{2\tilde{\gamma}}{(\tilde{\chi}-\tilde{\tau})^{3}} + 
\Phi'(m\tilde{\tau})
\right) \ln \frac{\tilde{\rho}^{2}}
{(\tilde{\chi}-\tilde{\tau})^{2}} \right.   \nonumber \\ 
& + &  \left. \frac{\tilde{\gamma}}
{(\tilde{\chi}-\tilde{\tau})^{3}} \Phi'(m\tilde{\tau})
\ln \frac{\tilde{\rho}^{2}}{\tilde{\tau}^{2}} + \ldots \right]
\label{Fin-1} \\
& = & - \frac{8 \alpha_{1}}{(192 \pi)^{2}} \frac{\tilde{\epsilon}^{3}}
{\Phi'(m\tilde{\tau})}
\left[ \ln \left( -\frac{\tilde{\epsilon}}{192 \pi^{2}
m\tilde{\tau} \Phi'(m \tilde{\tau})} \right) + \ldots \right].
  \label{Fin-2}
\eea
As we see from the last line, Eq. (\ref{Fin-2}),
the singular terms $\ln (1/\nu)$ cancel each other. 
Eq. (\ref{Fin-1}), reveals the reason  of this cancellation: due to
relation (\ref{saddle-eql}) the  coefficient of the term
$\ln (\tilde{\rho}^{2}/(\tilde{\chi}-\tilde{\tau})^{2})$, which gives rise to
the singularity $\ln (1/\nu)$, is equal to zero exactly. This result
is general and does not depend on any approximation.

We would like to remark that the terms singular in $\nu$ are proportional
to $\alpha_{1}$. From Eq. (\ref{R}) it follows that they originate from the
terms proportional to $s \ln s$ and $s$ in the residue of the instanton
propagator. If one uses the instanton propagator $G_{f}(p,q)$ satisfying
constraint (\ref{G-constr}) with the functions $f_{A}$ such that 
two leading terms in the asymptotics (\ref{G-asymp}) vanish, 
then the leading asymptotics $s\ln s$ and $s$ of the propagator for
large $s$ are absent. As a consequence, the singular terms $\ln (1/\nu)$
do not appear.

For energies small enough, such that $m \tilde{\tau} \ll 1$,
the expressions simplify further and the result for the next-to-leading
correction can be written in a simple form. We obtain that 
\beq
F^{(2)}(\epsilon,\nu) = - \frac{\alpha_{2}}{192 \pi^{2}} 
(\tilde{\epsilon}m \tilde{\tau})^{2} \left[ 
\ln \frac{\tilde{\epsilon} m \tilde{\tau}}{384 \pi^{2}} + \frac{58}{15} 
+ {\cal O}(m^{2}\tilde{\tau}^{2}) \right]. 
   \label{F2}
\eeq
In the limit $\epsilon \rightarrow 0$ we use solution
(\ref{tau-1}) and obtain
\beq
F^{(2)}(\epsilon, \nu) = \frac{ 4 \alpha_{2} \kappa^{2} \epsilon^{2}}
{192  \pi^{2}} \left( 1 + \frac{1}{2}
 \frac{\ln \ln \frac{1}{\epsilon}}{\ln \frac{1}{\epsilon}}
 + \ldots \right) =
\frac{\kappa^{2} \epsilon^{2}}{32  \pi^{2}}
 \left( 1 + \frac{1}{2}
 \frac{\ln \ln \frac{1}{\epsilon}}{\ln \frac{1}{\epsilon}}
 + \ldots \right)
\label{F2-1}
\eeq
We see that at low energies the main contribution is proportional to
$\alpha_{2}$, i.e. comes from the $\ln s$ and constant terms in the residue
(\ref{R}) of the  instanton propagator. In fact it is easy to check that
it is precisely the term $\sim \ln s$ in Eq. (\ref{R}) 
which gives the contribution (\ref{F2-1}).

\subsection{Numerical results}

For arbitrary $\epsilon$ and $\nu$ the 
functions $F^{(1)}(\epsilon,\nu)$ and $F^{(2)}(\epsilon,\nu)$ were 
studied numerically. It turned out that the saddle point solution 
exists only for a certain region in the $(\epsilon,\nu)$-plane. It 
lies inside the rectangle $0 < \epsilon < \epsilon_{max}= 0.55$ and 
$0 < \nu < \nu_{max} = 0.25$. We performed the numerical analysis for 
this whole region. 

To present the results it is convenient to introduce the following 
functions: 
\[
{\cal F}_{1}(\epsilon,\nu) = 1 - \frac{F^{(1)}(\epsilon,\nu)}{32 \pi^{2}},
{\cal F}_{2}(\epsilon,\nu) = 1 - \frac{F^{(1)}(\epsilon,\nu) + 
F^{(2)}(\epsilon,\nu)}{32 \pi^{2}}.
\]
They are normalized by the conditions 
${\cal F}_{1}(0,\nu) = {\cal F}_{2}(0,\nu) = 1$. 

\begin{figure}[ht]
\epsfxsize=0.9\hsize
\epsfbox{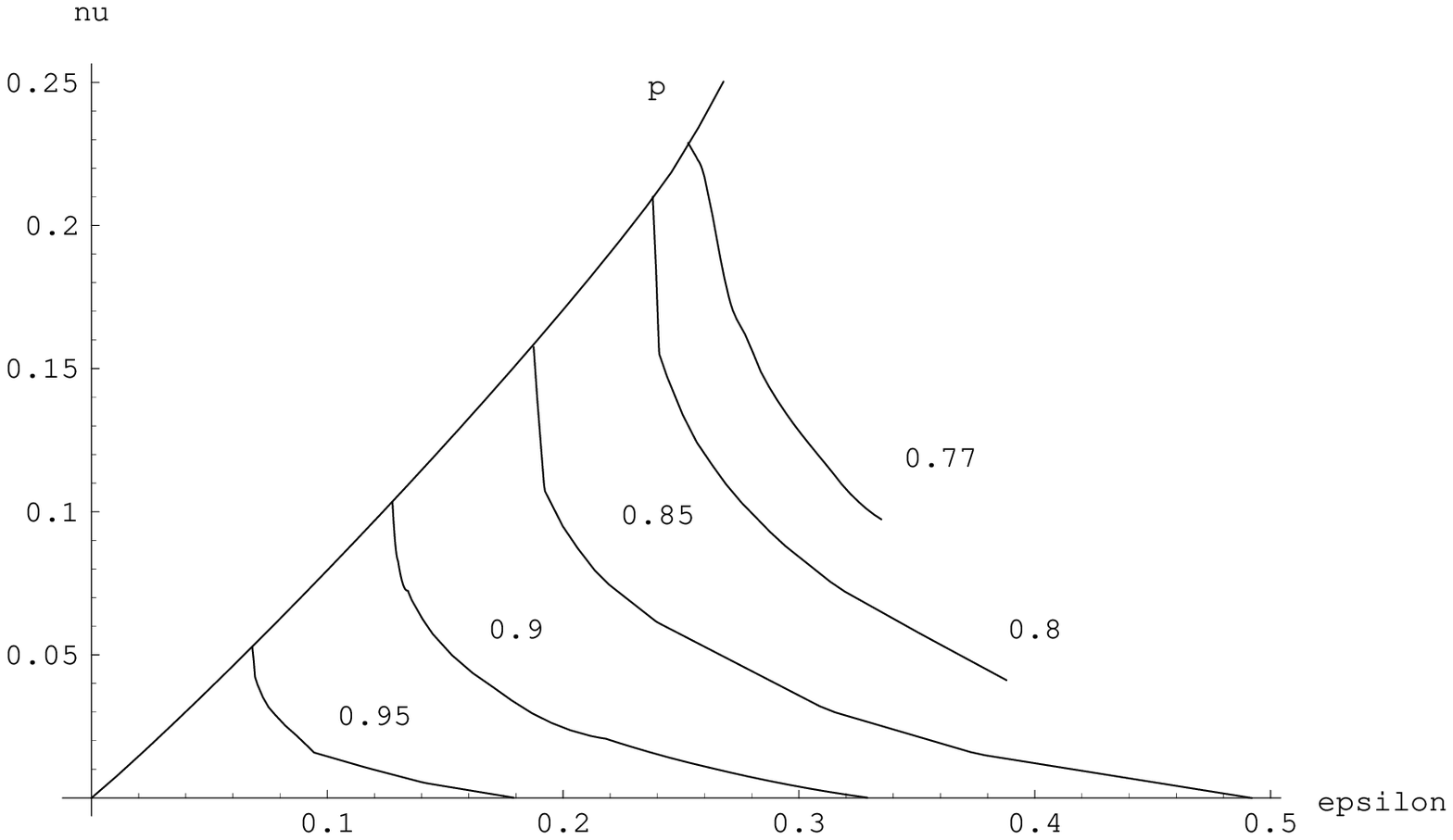}
%\centerline{\psfig{figure=inst-fig1.ps,height=12cm,width=16cm}}
%\centerline{\psfig{figure=inst-fig1.ps,height=12cm,width=8cm}}
\caption{Lines of constant ${\cal F}_{2}(\epsilon,\nu)$ in the 
$(\epsilon,\nu)$-plane. Numbers near the lines indicate the 
value of ${\cal F}_{2}$, ``p'' labels the line of periodic instanton 
solutions.}
\end{figure}

Lines of constant ${\cal F}_{2}(\epsilon,\nu)$ are plotted in Fig. 1. 
We want to study the cross section for shadow processes with a 
few initial particles. Then according to conjecture (\ref{conj}) 
points where the lines cross the $\nu = 0$ axis are of particular 
interest. For example, ${\cal F}_{2}(\epsilon,0)=0.95$ at 
$\epsilon = 0.180$, ${\cal F}_{2}(\epsilon,0)=0.85$ at 
$\epsilon = 0.492$. We would like to mention that, in fact, in the 
studied region of $(\epsilon,\nu)$ the propagator correction is 
quite small comparing to the leading order. Thus, the difference 
between ${\cal F}_{1}(\epsilon,\nu)$ and   
${\cal F}_{2}(\epsilon,\nu)$ does not exceed $10^{-2}$. 

The curves in Fig. 1 end at the line formed by saddle points 
corresponding to the periodic instanton solutions. For them 
$\tilde{\tau}(\epsilon,\nu) = \tilde{\chi}(\epsilon,\nu)/2$. This line 
is directed from the zero energy instanton ($\epsilon = \nu = 0$) to 
the sphaleron ($\epsilon = \nu =1$).  

As it has been already mentioned, the complete function 
$F(\epsilon,\nu)$ was calculated numerically in the range 
$0.4 < \epsilon < 3.5$ and $0.25 < \nu < 1$  in Ref. \cite{KT}. 
The computation was performed by solving a certain classical 
boundary value problem on the lattice. With the size of the 
lattice used in the numerical calculation in Ref. \cite{KT}, 
the authors did not obtain data for 
smaller $\epsilon$ and $\nu$ except for the line of the periodic 
instantons. The comparison shows that our perturbative results 
do not differ significantly from the exact ones of Ref. \cite{KT} 
for $\epsilon < 0.25$ and $\nu < 0.2$. These values can be regarded as a 
rough estimate of the range of validity of the leading and next-to-leading 
approximations.     

\section{Discussion and conclusions}

In the present paper we have analyzed the multiparticle cross section
of the shadow processes induced by instanton transitions in the simple
scalar model (\ref{action}). Using the exact analytical
expression for the on-shell residue of the propagator of quantum
fluctuations in the instanton background we calculated 
the suppression factor in the next-to-leading order.

The calculation of the leading and next-to-leading orders 
of $F(\epsilon,\nu)$ was performed assuming that the size of the 
instanton solution is small enough, namely $(\tilde{\rho}m) \ll 1$. 
Neglecting ${\cal O}(\rho^{4} m^{4})$ terms in the action (\ref{S-ci}) 
and using the instanton and the residue of the instanton propagator 
of the {\it massless} theory in Eqs. (\ref{R-inst-def}) and (\ref{RR1}) 
amount to omission of corrections of the type 
\beq
  \frac{\rho^{2}}{\tau^{2}} (\rho^{2} m^{2})  \label{corr1}
\eeq
in $F^{(1)}$ and 
\beq
(\rho m)^{4}, \; \; \frac{\rho^{4}}{\tau^{4}} (\rho^{2} m^{2}), \; \; 
\frac{\rho^{4}}{\tau^{4}} (\rho^{2} m^{2}) \ln \frac{\rho^{2}}{\tau^{2}}
\; \; \mbox{and} \; \; 
 \frac{\rho^{4}}{\tau^{4}} (m^{2}\tau^{2})^{k}   \label{corr2}
\eeq
in $F^{(2)}$. We checked numerically that in the region of 
$\epsilon$ and $\nu$, where the saddle point solution exists, the  
terms in Eqs.(\ref{corr1}), (\ref{corr2}) are really small. As an 
illustration let us consider the case of very small $\epsilon$ 
and use the saddle point solution (\ref{tau-1}), (\ref{saddle-1}). 
We obtain that 
\bea
\frac{\tilde{\rho}^{2}}{\tilde{\tau}^{2}} (\tilde{\rho}^{2} m^{2}) & \sim & 
\frac{\epsilon^{2}}{\left(\ln \frac{1}{\epsilon} \right)^{5/2}}, \nonumber \\
(\tilde{\rho} m)^{4} & \sim & 
\frac{\epsilon^{2}}{\left(\ln \frac{1}{\epsilon}\right)^{3}}, \; \; \; \; 
\frac{\tilde{\rho}^{4}}{\tilde{\tau}^{4}} (\tilde{\rho}^{2} m^{2}) \sim 
\frac{\epsilon^{3}}{\left(\ln \frac{1}{\epsilon}\right)^{5/2}}, \nonumber \\ 
\frac{\tilde{\rho}^{4}}{\tilde{\tau}^{4}} (\tilde{\rho}^{2} m^{2}) 
\ln \frac{\tilde{\rho}^{2}}{\tilde{\tau}^{2}} 
& \sim & \frac{\epsilon^{3}}{\left(\ln \frac{1}{\epsilon}\right)^{3/2}}, 
\; \;  \; \; 
 \frac{\tilde{\rho}^{4}}{\tilde{\tau}^{4}} (m^{2}\tilde{\tau}^{2})^{k} \sim 
\frac{\epsilon^{2}}{\left(\ln \frac{1}{\epsilon}\right)^{1+k/2}}.  
\nonumber 
\eea
All these corrections are subleading compared to the terms 
retained in the function $F^{(2)}(\epsilon,\nu)$, Eq. (\ref{F2-1}). 
Contributions due to non-zero mass amount to corrections in powers of 
$(m \tilde{\tau})$, where $\tilde{\tau}(\epsilon,\nu)$ is the saddle point 
solution for $\tau$. In general, these corrections are not small, and  
all of them were taken into account by using the $s$-variable and 
the energy $\omega_{\bk}$ of  massive particles in Eqs. 
(\ref{RR1}) - (\ref{sab}), (\ref{W1}), (\ref{I-def}), etc. 
Our numerical analysis shows that the inequality 
$m\tilde{\tau} <1$ is satisfied, for example, for $\epsilon < 0.4$ if 
$\nu$ is close to $\nu =0$ and for $\epsilon < 0.02$ for the periodic 
instanton solutions. Comparing this to the region in the 
$(\epsilon,\nu)$-plane in Fig. 1, for which we carried out 
the calculation in this article, one can see that our formalism, 
accounting for arbitrary $m\tilde{\tau}$, allows to 
enlarge considerably the range of validity 
of the next-to-leading approximation.  

The range of validity of the next-to-leading order approximation 
of the function $E(\epsilon,\nu)$ was estimated by
comparing our results with numerical computations 
in Ref. \cite{KT} for the values of $\epsilon$ and $\nu$ for which the
latter can be translated to the case of shadow processes, i.e., for
periodic instantons.  The comparison shows that the perturbative
results do not differ significantly from the exact ones for
$\epsilon \leq 0.25$ and $\nu \leq 0.2$.  

For this range of values of $\epsilon$ and $\nu$ and away from the
line of periodic instantons, methods of Ref. \cite{KT} do not allow
to obtain exact results. Therefore, at the moment our perturbative 
calculations are the only ones which give quantitative behaviour of the
supression factor in this range.

{}From Eqs. (\ref{tau-1}) we see that approximate formulas
(\ref{F1}) and (\ref{F2-1}) are valid as long as
\[
\frac{\ln \ln \frac{1}{\epsilon}}{\ln \frac{1}{\epsilon}} \ll 1.
\]
For this range of energies we obtained the analytical expressions for 
the suppression factor and values of the saddle point parameters 
$\tilde{\rho}$, $\tilde{\chi}$, $\tilde{\tau}$ and $\tilde{\gamma}$. 
Formula (\ref{F2}) for the propagator correction for small $\nu$ is 
valid when $m \tilde{\tau} \ll 1$. According to the estimate, 
mentioned above, this condition is satisfied if $\epsilon \ll 0.4$. 
This can be also verified by analyzing Eq. (\ref{tau-eqn}).  

We also checked the cancellation of terms singular in the limit
$\nu \rightarrow 0$ in the propagator correction $F^{(2)}$.
As we have explained, this is closely related to the problem
of quasiclassical evaluation of contributions of initial states and
initial-final states. In the article we also discussed this problem
within the approach proposed by Mueller. Namely, we calculated
the leading asymptotics of the instanton propagator at large $s$ and
showed that it can be cancelled by an appropriate choice of the
propagator constraint. According to Ref. \cite{Mu92}, with such 
propagator the problem of semiclassical calculation of contributions 
due to initial states and initial-final states can be tackled properly. 

\section*{Acknowledgments}

{\tolerance=500
We would like to thank A. Ringwald and V. Rubakov for
discussions and valuable comments.  Y.K. acknowledges financial
support from the Russian Foundation for Basic Research (grant
98-02-16769-a) and grant CERN/P/FIS/1203/98.


\begin{thebibliography}{99}

\bibitem{shadow} M.B. Voloshin, {\em Nucl. Phys.} {\bf B363} (1991) 425. \\
S.D.H. Hsu, {\em Phys. Lett.} {\bf B261} (1991) 81. 

\bibitem{EWT-vac} 
C.G. Callan, R.F. Dashen, and D.J. Gross, {\em Physl Lett.} {\bf 63B} 
(1976) 334.

\bibitem{VolKobOkun} 
M.B. Voloshin, I.Yu. Kobzarev, and L.B. Okun, {\em Sov. J. Nucl. Phys.} 
{\bf 20} (1975) 644. 

\bibitem{Ri1}
A. Ringwald, {\em Nucl. Phys.} {\bf B330} (1990) 1. \\
O. Espinosa, {\em Nucl. Phys.} {\bf B334} (1990) 310.

\bibitem{KRT}
S.Yu. Khlebnikov, V.A. Rubakov and P.G. Tinyakov,
{\em Nucl. Phys.} {\bf B350} (1990) 441.

\bibitem{Mat}
M.~Mattis, {\em Phys. Rep.} {\bf 214} (1992) 159.

\bibitem{Ti1}
P.G. Tinyakov, {\em Int. J. Mod. Phys.} {\bf A8} (1993) 1823.

\bibitem{KT}
A.N. Kuznetsov and P.G. Tinyakov, {\em Phys. Rev.}
{\bf D56} (1997) 1156.

\bibitem{Mu92}
A.H. Mueller, {\em Nucl. Phys.} {\bf B381} (1992) 597.

\bibitem{RT}  
V. A. Rubakov and P.G. Tinyakov,
{\em Phys.Lett.} {\bf B279}, 165 (1992).

\bibitem{Ti2}
P.G. Tinyakov, {\em Phys. Lett.} {\bf B284} (1992) 410.

\bibitem{RST} 
V.A. Rubakov, D.T. Son, and P.G. Tinyakov, {\em Phys.Lett.} 
{\bf B287} (1992) 342. 

\bibitem{Mueller} 
A.H. Mueller, {\em Nucl.Phys.} {\bf B401} (1993) 93. 

\bibitem{RubReb}
G.F. Bonini, A.G. Cohen, C. Rebbi, and V.A. Rubakov, 
{\em Phys. Rev.} {\bf D60} (1999) 076004; 
e-Print Archive: hep-ph/9901226. 

\bibitem{Fu}
S. Fubini, {\em Nuovo Cimento} {\bf 34A} (1976) 521.

\bibitem{Lip}
L.N. Lipatov, {~\em Zh. Eksp. Teor. Fiz.} {~\bf 72}
(1977) 411; {~\em Sov. Phys. JETP} {~\bf 45} (1977) 216.

\bibitem{Aff}
I. Affleck, {\em Nucl. Phys.} {\bf B191} (1981) 429.

\bibitem{LevYaf}
H. Levine and L.G. Yaffe, {\em Phys. Rev.} {\bf D19} (1979) 1225.

\bibitem{MMcLY}
M. Mattis, L. McLerran, and L. Yaffe, {\em Phys.Rev.} {\bf D45} (1992)
4294.

\bibitem{Ku}
Yu.A. Kubyshin, {~\em Teor. Mat. Fiz.} {~\bf 57}
(1983) 363; {\em Theor. Math. Phys.} {\bf 57} (1984) 1196.

\bibitem{KubTi}
Yu.A. Kubyshin and P.T. Tinyakov, hep-ph/9812321.

\bibitem{Vo1}
M.B.Voloshin, {\em Nucl. Phys.} {\bf B363} (1991) 425.

\end{thebibliography}
\end{document}